\documentstyle[aps,multicol,amsfonts,graphicx]{revtex}
\newcommand{\fs}{\footnotesize}
\begin{document}
\setcounter{page}{243}
\noindent
{\sc SOVIET PHYSICS JETP\hfill VOLUME 30, NUMBER 2\hfill FEBRUARY, 1970}\\[10mm]
{\sl RESONANT RADIATIVE PROCESSES }
\\[2mm]
{ T.Ya.POPOVA, A.K.POPOV, S.G.RAUTIAN, and A.A.FEOKTISTOV}
\\[2mm]
{\small Semiconductor Physics Institute, Siberian Division,
U.S.S.R. Academy of Sciences}
\\[1ex]
{\small Submitted December 20,1968 }
\\[1ex]
{\small Zh. Eksp. Teor. Fiz. 57, 444-451 (August, 1969)}
\\
\small
\begin{abstract} The frequency correlation properties of the radiation from an atom
in a strong field in resonance with neighboring transitions are considered.
It is shown that the difference in frequency correlation in two-photon and
stepwise processes decreases with increase of the external field. The spectral
compositions of the Doppler-broadened resonance scattering and fluorescence
are analyzed. It is shown that in these cases the Doppler line width is
anisotropic.
\end{abstract}

\begin{multicols}{2}
\narrowtext
\section{INTRODUCTION}

\par
Thermal motion of radiating atoms leads, owing to the Doppler effect, to an isotropic
broadening of the luminescence line. At the same time, the line width of Rayleigh
scattering depends on the direction [1]

{\fs$$ \Delta \omega= \omega \frac{\bar v}{c} \cdot 2\sin\frac{\theta}{2}, \qquad
{\bar v}={\sqrt {2kT/m}}, $$}

and vanishes in the case of forward scattering ($\theta = 0$). This difference in the
manifestation of the Doppler effect is due entirely to the deference between the
frequency-correlation properties of the indicated processes [2]. On the other hand,
the frequency-correlation properties are strongly pronounced also in other two-photon
processes (two-quantum absorption and luminescence, or Raman scattering). It is
thereforenatural to expect the existence of anisotropy of the Doppler line width in
this case, too. This question is discussed in Sec. 3.

We note now that an analysis of radiative processes within the framework
of second order perturbation-theory leads to a delimitation of two-photon
processes proper from stepwise (or cascade) processes, for example two-photon
luminescence and cascade emission of two photons with a real intermediate state.
Such a delimitation is based essentially on the frequency-correlation
properties. On the other hand, if the energy of interaction of the atom with
the field is larger than the level width, then the frequency-correlation
properties of the radiative processes experience a strong metamorphosis
(Sec. 2). In particular, two-photon and stepwise processes turn out to be
physically Indistinguishable. As a consequence, in strong fields the
manifestation of the Doppler broadening also changes strongly. The resultant
phenomena are traced for the resonant-scattering doublet and
resonant-fluorescence triplet (Secs. 4 and 5).

\section {FREQUENCY -- CORRELATION PROPERTIES OF RADIATIVE PROCESSES}
\par
We shall consider the radiation of an atom situated in an external field with two
monochromatic components of frequencies $\omega$ and $\omega_{\mu}$ and amplitudes
$E$ and $E_{\mu}$. We assume that each field component interacts only with one
transition (resonance approximation). In the scheme corresponding to Raman scattering
(Fig. 1), a photon  $\hbar \omega$ is absorbed and a photon $\hbar \omega_{\mu}$ is
emitted. The probability amplitudes $a_i$ of the states $i = m, n, l$ satisfy the
system of equations
\begin{figure}[t]
\[\includegraphics[width=0.2\textwidth]{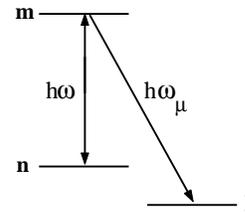}\]
\vspace*{-0.5cm}\caption{\small Term scheme.}\label{fig1} \end{figure}

{\fs $$ \dot a_m+ \gamma_m a_m= iGe^{-i\Omega t} a_n + iG_{\mu}e^{-i\Omega_{\mu} t}
a_l, $$ $$ \dot a_n+ \gamma_n a_n= iG^{*}e^{i\Omega t} a_m, \quad \dot a_l+ \gamma_l
a_l= iG^{*}_{\mu}e^{i\Omega_{\mu} t} a_m, \eqno{(2.1)} $$ $$
\Omega=\omega-\omega_{mn},\ \Omega_{\mu}=\omega_{\mu}-\omega_{ml},\
G=d_{mn}E/2\hbar,\ G_{\mu}=d_{ml}E_{\mu}/2\hbar, $$} where $d_{ij}$ are the matrix
elements of the dipole moment.

We shall henceforth regard $G e^{-i\Omega t}$ as a strong perturbation and
$G_{\mu}e^{-i\Omega_{\mu}t}$ as a weak perturbation. We are interested in the
probability $ w_{\mu}$ of emission of the photon $\hbar \omega_{\mu}$. The solution
of the system (2.1) can be obtained by successive approximations in the parameter
$G_{\mu}$ [3-5]: we consider the system of equations

{\fs$$
\dot a_m+ \gamma_m a_m= iGe^{i\Omega t} a_n, \quad
\dot a_n+ \gamma_n a_n= iG^{*}e^{i\Omega t} a_m,        \eqno{(2.2)}
$$}

in the zeroth approximation and introduce Its exact solution into the right-hand
side of the equation for $a_l(t)$ in (2.1). Integration of this equation yields
the first approximation $G_{\mu}$ for $a_l(t)$, with the aid of which we
calculate

{\fs$$
w_{\mu}=2\gamma_l \int_0^{\infty} {|a_l(t)|^2dt}.
$$}

The solution of the system (2.2) can be represented in the form

{\fs$$ a_n(t)=[A_1e^{-\alpha_1 t}+A_2e^{-\alpha_2 t}]\, e^{i\Omega t},$$ $$a_m(t)= iG
\left [\frac {A_1e^{-\alpha_1 t}}{\gamma_m - \alpha_1}+ \frac {A_2e^{-\alpha_2
t}}{\gamma_m - \alpha_2} \right ], \eqno {(2.3)} $$ $$ \alpha_{1,2}=\frac {\Gamma + i
\Omega}{2} \pm i \sqrt {G^2+ \left ( \frac {\Omega-i\gamma}{2} \right )^2}, \ \Gamma
= \gamma_m+ \gamma_n, \ \gamma=  \gamma_n- \gamma_m, \eqno {(2.4)} $$} where
$A_{1,2}$ are the integration constants. In the case of interest to us, $a_n(0)=1$,
we have

{\fs$$ A_{1,2}=\mp \frac {\gamma_m-\alpha_{1,2}}{\alpha_1-\alpha_2}, \quad a_m(t)=
\frac {iG}{\alpha_1-\alpha_2}\,[\,e^{-\alpha_1 t}-e^{-\alpha_2 t}], \eqno(2.5) $$ $$
w_{\mu}= \frac {2|GG_{\mu}|^2}{|\alpha_1-\alpha_2|^2} {\rm Re} \left \{ {\frac
{(\alpha_1+\alpha_1^*)^{-1}-(\alpha_2+\alpha_1^*)^{-1}}
{\gamma_l+\alpha_1^*+i\Omega_{\mu}}}+\right.$$ $$\left.+ \frac
{(\alpha_2+\alpha_2^*)^{-1}-(\alpha_1+\alpha_2^*)^{-1}}
{\gamma_l+\alpha_2^*+i\Omega_{\mu}} \right \}. \eqno{(2.6)} $$}

The frequency-correlation properties consist in the fact that the frequencies
$\Omega_{\mu}$ at which there is maximum probability of emitting the photon
$\hbar \omega_{\mu}$ ($\Omega_{\mu1}={\rm Im} \, \alpha_1,
\Omega_{\mu2}={\rm Im} \, \alpha_2$) turn out to depend on $\Omega$, i.e.,
on the frequency of the absorbed photon.
In the case of small $G$, we have in place of (2.4) and (2.6)

{\fs$$ \alpha_1=\gamma_m, \quad \alpha_2=\gamma_n+i\Omega, \quad G\ll|\Omega-i\gamma|
$$ $$ w_{\mu}= \frac {|GG_{\mu}|^2}{|\gamma_n-\gamma_m+i\Omega|^2}\times$$
$$\times{\rm Re} \left \{ \frac
{\gamma_m^{-1}-2(\gamma_m+\gamma_n+i\Omega)^{-1}}{\gamma_l+\gamma_m+i\Omega_{\mu}}+
\frac
{\gamma_n^{-1}-2(\gamma_m+\gamma_n-i\Omega)^{-1}}{\gamma_l+\gamma_n+i(\Omega_{\mu}-\Omega)}
\right \}. \eqno(2.7) $$}

The second term in the expression for $w_{\mu}$ reaches a maximum at a frequency
$\omega_{\mu}=\omega-\omega_{ln}$ (Raman scattering). It can therefore be said that in
the second stage of the two-process the atom "remembers" which quantum was absorbed
during the first stage. The Raman-scattering line width $\gamma_l+\gamma_n$ also
"remembers" from which level the atom arrived at the first stage. These indeed are
the properties of frequency correlation. To the contrary, the first term in (2.7)
gives resonance at $\Omega_{\mu}=\omega_{\mu}-\omega_{ml}=0$, i.e., at the frequency
of the transition between the intermediate and final states of the atom. The width of
the corresponding line is also determined by the levels $m$ and $l$. This term
describes the cascade or stepwise transition $n \rightarrow m \rightarrow l$, and
there is no correlation in it at all between the absorption and emission acts.

The correlation properties of the emission processes are closely connected with the
type of evolution of amplitude $a_m(t)$ of the intermediate state. Let $|\Omega|\gg
\gamma_m,\gamma_n$; then the rapidly oscillating term $\exp[-(i\Omega+\gamma_n)t]$
(virtual state) carries information concerning the initial state ($\gamma_n$) and the
absorbed quantum ($\Omega$), and causes the appearance of a scattering line, i.e.,
the second term in (2.7) (the terms $2(\gamma_m+\gamma_n \pm i\Omega)^{-1}$ can be
discarded). The time dependence of the second term, $exp(-\gamma_m t)$, contains no
attributes of the absorption act and does not differ from the case when the state $m$
is the initial state (i.e., $a_m(0) = 1$). We can therefore say with respect to this
term and with respect to the corresponding unshifted line at the transition $m
\rightarrow l$ that the intermediate state is a real state of the atom, having a
finite lifetime $(2\gamma_m)^{-1}$.

A separate examination of the transitions through the virtual and real states
signifies that only squares of the moduli of the first and second terms remain in the
expression for $|a_l(t)|^2$. The crossing term lead to the appearance of $2(\gamma_m
+ \gamma_n \pm i \Omega)^{-1}$ in ${w}_{\mu}$, and it is legitimate to neglect them
if $|\Omega| \gg \gamma_m, \gamma_n$. In the case when $|\Omega| \sim
\gamma_m,\gamma_n$, the crossing terms, which reflect the interference between the
real and virtual states, are significant and cannot be discarded. However, even here
${w}_{\mu}$ can be represented in the form of a sum of terms with and without
"memory", and allowance for the interference only changes the coefficient preceding
these terms. The only physical basis for contrasting the stepwise and two-photon
processes is the difference between their frequency-correlation properties, which are
uniquely connected with the singularities of the evolutions of the individual terms
of the amplitude $a_m(t)$ of the intermediate state. On the other hand, formulas
(2.7) are valid within the framework of second-order perturbation theory, and are no
longer valid at sufficiently large $G^2$. In the general case, $a_m$ contains two
exponential terms (formula (2.5)), which are formally analogous to the "virtual" and
"real" states. However both $\alpha_1$ and $\alpha_1$ depend on the parameters of the
field $(G^2, \Omega)$ and of the two combining levels $(\gamma_m, \gamma_n)$. With
respect to ${w}_{\mu}$, this means that in both resonances the atom "remembers" which
quantum was observed during the first stage of the process. In the limiting case of a
very strong field we have

{\fs$$ \alpha_{1,2} \simeq [\gamma_m+\gamma_n+ i\Omega]/2 \pm iG, \quad G \gg
|\Omega- i \gamma|, \eqno(2.8) $$}

i.e., the differences in the temporal properties of the two exponentials in (2.5)
have disappeared completely. At the same time, the differences in the
frequency-correlation properties of the corresponding lines have also disappeared,
and there are no grounds for distinguishing between the two. The concepts of stepwise
and two-photon transitions or of the virtual and real states are likewise physically
Indistinguishable. It is clear from the foregoing that these concepts are inseparably
linked with perturbation theory and lose physical meaning outside the region of its
applicability.

The external field levels out the differences between $\alpha_1$ and $\alpha_2$
both with
respect to the yield of the resonance $(\Omega)$ and with respect to the difference
In the damping of the states $m, n$. Let us consider in greater detail
the case $\gamma_m=\gamma_n=\Gamma$, when the leveling function of the field
simplifies:

{\fs$$ \alpha_{1,2}=\Gamma+i \alpha_{1,2}^{''}, \quad \alpha_{1,2}^{''}= \frac{1}{2}
[\Omega \pm \sqrt {\Omega^2+4G^2}]. \eqno(2.9) $$}

\begin{figure}[b]
\[\includegraphics[width=0.3\textwidth]{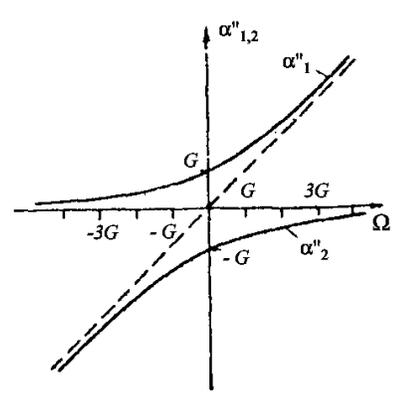}\]
\vspace*{-0.5cm}\caption{\small Plot of $\alpha''_{1,2}$ against
$\Omega$.}\label{fig2} \end{figure}

Figure 2 shows plots of $\alpha_{1,2}^{''}$ as functions of $\Omega$. The asymptotic
approach of the plots to the abscissa axis and to the dashed line
$\alpha^{''}=\Omega$ corresponds respectively to real and virtual states. As $\Omega$
changes from positive to negative values, we have for $\alpha_{1}^{''}$, for example,
a smooth transition from the properties of the virtual state to the properties of the
real state, and for $\alpha_{2}^{''}$ the inverse sequence. In the region
$|\Omega|<G$, the states $\alpha_1$ and $\alpha_2$ differ little.

As the measure of the "memory" of the absorbed quantum we can choose the
quantity

{\fs$$ M_{1,2} \equiv \frac {d \alpha_{1,2}^{''}}{d \Omega}= \frac {1}{2} \left [ 1
\pm \frac {\Omega}{\sqrt {\Omega^2+4G^2}} \right ], \eqno(2.10) $$} which varies from
0 to 1. The value $M = 0$ means complete absence of memory (stepwise transition),
while $M = 1$ corresponds to total correlation between the frequencies of the
absorbed and emitted photons (two-photon transition). Since
$\alpha_1+\alpha_2=\Gamma+i\Omega$, it follows that $M_1+M_2 = 1$; this fact can be
interpreted as follows: the intermediate state as a whole, without subdivision into
two states, retains the entire information on the absorbed photon. In the case
$|\Omega| \ll G$, we have $M_{1,2} \approx 1/2$, i.e., the "memory" is equally
divided between the two terms in $a_m(t)$.

If $\alpha_{1}^{''}-\alpha_{2}^{''}$ is sufficiently small, the interference of the
states $\alpha_1$ and $\alpha_1$ is quite significant. A separate analysis of the
transitions through these intermediate states is meaningful if the distances between
the resonances exceed their widths:

{\fs$$
|\alpha_{1}^{''}-\alpha_{2}^{''}| \gg \alpha_{1}^{'}+\alpha_{2}^{'}+2\gamma_l.
$$}

Under these conditions, the emission spectrum for the transition $m \rightarrow l$
has the form of a well resolved doublet, the appearance of which can be interpreted
as the splitting of the intermediate-state level into two sublevels [3-9]. In
accordance with the foregoing, It is meaningless to attribute the components of this
doublet to stepwise or two-photon transitions. One can only speak of the
resonance-scattering doublet as a unit. We shall henceforth use this term. {\it For
concreteness}, we have referred throughout to processes of the type of Raman
scattering. {\it All the physical conclusions pertain also to other processes in which
two-photons take part}, such as two-quantum luminescence, {\it two-quantum
absorption}, and Raman scattering via a lower intermediate level. {\it It is only
necessary to reverse the signs of\, $\Omega$ and $\Omega_{\mu}$ in all the formulas,
depending on whether the corresponding guantum is emitted or absorbed}.

\section{DOPPLER  BROADENING  OF  RAMAN  SCATTERING LINE}

\par
Allowance for the motion of the atom in the case of traveling waves reduces, as is
well known, to the substitutions $\Omega \rightarrow \Omega-{\bf k \cdot v}$ and
$\Omega_{\mu} \rightarrow \Omega_{\mu}-{\bf k}_{\mu} {\bf \cdot v}$, where ${\bf k}$
and ${\bf k}_{\mu}$ are the wave vectors of the waves. This means that the amplitudes
$A_1$ and $A_2$ also depend on the velocity of the atom. Within the framework of the
second approximation of perturbation theory, this pertains only to the virtual
sublevel. This case apparently {\it has not been discussed in the literature}, and
will be considered in the present section. Averaging over the velocities will be
carried with a Maxwellian distribution:

{\fs$$ (\pi {\rm\bar v})^{-3/2} \exp \{ -{\rm\bf v}^2/ {{\rm\bar v}^2}\}. \eqno (3.1)
$$}

Let the deviation from resonance be larger not only than the natural width
but also of the Doppler width $(|\Omega| \gg {\bf k {\rm\bf\bar v}})$.
Under this condition we can neglect
the interference between the real and virtual states, and we can easily obtain
from (2.7)

{\fs$$ \langle w_{\mu} \rangle= \frac {|GG_{\mu}|^2}{\Omega^2} {\rm Re} \left \{
\frac {\sqrt \pi}{\gamma_m k_{\mu} {\rm\bar v}}\,e^{p_1^2}[1-\Phi (p_1)]+ \frac {\sqrt
\pi}{\gamma_n  q{\rm\bar v}}\,e^{p_2^2}[1+\Phi (p_2)] \right \}, \eqno(3.2) $$ $$
p_1=[\gamma_l+\gamma_m+i\Omega_{\mu}]/k_{\mu} {\rm\bar v}, \quad
p_2=[\gamma_l+\gamma_n+i(\Omega_{\mu}-\Omega)]/q \rm\bar v, $$ $$ q=|{\bf
k}_{\mu}-{\bf k}|= \sqrt {(k_{\mu}-k)^2+4k_{\mu}k\sin^2(\theta/2)}, $$} \noindent
where $\Phi (z)$ is the probability integral and $\theta$ is the angle between ${\bf
k}$ and ${\bf k_{\mu}}$. If the Doppler broadening dominates over the natural
broadening, $k_\mu {\rm\bar v} \gg \gamma_l+\gamma_m,\, q {\rm\bar v} \gg
\gamma_l+\gamma_n$, then we can assume that $\Phi (p_{1,2}) = 0$ and (3.2) contains
two terms of Gaussian form

{\fs$$ \langle w_{\mu} \rangle= \frac {|GG_{\mu}|^2}{\Omega^2} \left \{ \frac {\sqrt
\pi}{\gamma_m k_{\mu} {\rm\bar v}} \exp \left [- \frac {\Omega_\mu^2}{(k_\mu {\rm\bar
v})^2} \right ]+\right.$$ $$\left.+ \frac {\sqrt \pi}{\gamma_n q {\rm\bar v}} \exp
\left [- \frac {(\Omega_\mu-\Omega)^2}{( q{\rm\bar v})^2} \right ] \right \}.
\eqno(3.3) $$}

The first terms in (3.2) or (3.3) (stepwise-transition line) has a Doppler width
$k_\mu \rm\bar v$, which does not depend on $\theta$. On the other hand, the Doppler
width $q \rm\bar v$ of the Raman-scattering line {\it depends strongly on the
observation direction, changing from $|k_\mu - k| \rm\bar v$ to $(k_\mu + k) \rm\bar
v$} when $\theta$ changes from zero to $\pi$ (Fig. 3a). If $|k_\mu - k| \rm\bar v \ll
\gamma_l+\gamma_n$, then the Raman scattering {\it has a Lorentz shape} in the angle
interval $\theta \le  \gamma_l+\gamma_n)/k_\mu \rm\bar v$

{\fs$$
\frac {(\gamma_l+\gamma_n)/\gamma_n}{(\gamma_l+\gamma_n)^2+(\Omega_\mu+\Omega)^2}
$$}

\noindent {\it and its width is determined by the natural damping of the initial and
final states}. On the other hand, for the direction $\theta=\pi$, {\it the width $(2k
{\rm\bar v})$ is twice the width of the stepwise-transition line}.
\begin{figure}[b]
\[\includegraphics[width=0.25\textwidth]{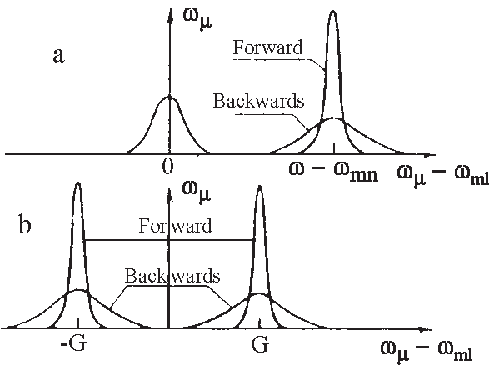}
\includegraphics[width=0.22\textwidth]{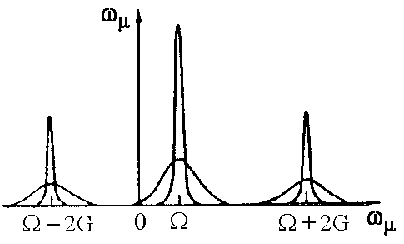}\]
{\caption{\small Spectrum of Raman scattering and stepwise transition (a) and
spectrum of Doppler-broadened resonant scattering (b).}}\label{fig3}{\caption{\small
Resonance fluorescence triplet.}}\label{fig4}
\end{figure}

It is easy to show that the doublet-component intensities integrated with respect to
$\Omega_\mu$ do not depend on $\theta$. Consequently, the width anisotropy means also
an angular dependence of the ratio of the intensities at the maxima of the lines in
the range $k_\mu \gamma_m/|k_\mu-k|\gamma_n$.

Formula (3.2) for $q{\rm\bar v}$ is analogous to the expression for the line width of
the Rayleigh scattering in a gas, $2k{\rm\bar v} \sin(\theta/2)$ [1], which is
obtained from (3.2) when $k = k_\mu$. Just as in the case of Rayleigh scattering,
formulas (3.2) and (3.3) admit of a simple interpretation, if we consider Raman
scattering as emission of a classical oscillator moving with velocity ${\rm\bar v}$.
A change-over to the c.m.s. of the oscillator changes the frequency $\omega$ of the
external field by $\omega-{\bf k} \cdot {\rm\bf v}$. The forced oscillations induced
by the field also have a frequency $\omega-{\bf k} \cdot {\rm\bf v}$. The internal
motion in the atom (or in the molecule) with natural frequency $\omega_{ln}$
modulates the forced oscillation and leads to the appearance in the emission spectrum
of a component of frequency $\omega-{\bf k} \cdot {\rm\bf v}-\omega_{ln}$. Finally,
for the wave emitted in the ${\bf k}_\mu$ direction, the inverse transition to the
stationary system of coordinates yields a frequency $\omega-\omega_{ln}-({\bf k}_\mu
- {\bf k}) \cdot {\rm\bf v}$, and averaging over $\rm\bf v$ leads to a Doppler width
$|{\bf k}_\mu - {\bf k}|{\rm\bar v}=q{\rm\bar v}$.

\section{DOPPLER \ BROADENING \ OF \ RESONANT -- SCATTERING DOUBLET}

\par
Let us turn now to the strong-field problem and assume that $G \gg k{\rm\bar v}$.
Here, too, we can neglect the "interference" terms $(\alpha_1+\alpha_2^*)^{-1}$ and
$(\alpha_2+\alpha_1^*)^{-1}$ in the expression (2.6) for ${w}_\mu$, and the Doppler
shifts of $\alpha_{1,2}$ can be taken into account in the first nonvanishing
approximation:

{\fs$$
\alpha_j=\alpha_{j0}-iM_j {\bf kv}, \quad j=1,2,
\eqno (4.1)
$$}

\noindent where $\alpha_{j0}$ are the values of $a_j$ at ${\bf k}  \cdot {\rm\bf v}=
0$ and $M_j$ is the "memory" factor determined by formula (2.10). Neglecting the
difference between $\alpha_{j}$ and $\alpha_{j0}$ everywhere except in the resonant
denominators, we get from (2.6)

{\fs$$ \langle w_\mu \rangle =\frac {|GG_\mu|^2}{\Omega^2+4G^2}{\rm Re} \left \langle
\frac {\gamma_m^{-1}}{\gamma_l+\alpha_{10}^*+i[\Omega_\mu-({\bf k}_\mu-M_1{\bf
k}){\bf v}]}\times\right.$$ $$\left. \times \frac
{\gamma_n^{-1}}{\gamma_l+\alpha_{20}^*+i[\Omega_\mu-({\bf k}_\mu-M_2{\bf k}){\bf v}]}
\right \rangle. \eqno(4.2) $$}

\noindent
 Averaging of this expression leads to a
formula of the type (3.2), in which the quantities $p_{1,2}$ should be replaced by

{\fs$$ p_j=[\gamma_l+\alpha_{j0}^*+i\Omega_\mu]/q_j {\rm\bar v}, \ q_j=\sqrt
{(k_\mu-M_j k)^2+4M_j k k_\mu \sin^2 (\theta/2)},$$ $$ \quad j=1,2, \eqno(4.3) $$}

\noindent and the widths $k_\mu {\rm\bar v}$ and $q{\rm\bar v}$ should be replaced by
$q_1{\rm\bar v}$ and $q_2{\rm\bar v}$. Thus, the widths and the positions of both
lines in the $m \rightarrow l$ transition depend on the frequency and direction of
propagation of the absorbed photon, and its role is determined by the "memory" factor
$M_{1,2}$. In the case of exact resonance we have $M_1=M_2=1/2$ and
$\alpha_{1,0}^{''}=-\alpha_{2,0}^{''}=|G|$, i.e., both lines are simmetrical relative
to the frequency $\omega_{ml}$ and have an identical angular dependence of the width
(Fig. 3b). {\it It is interesting that when the condition $k_\mu = M_1 k$ or $k_\mu =
M_2 k$ is satisfied, one of the lines has a natural width} in the case of observation
along $\theta = 0$ ({\it see the discussion of formulas (3.2) and (3.3)}).

Thus, variation of $\Omega$ leads to the following changes in the spectrum. When
$|\Omega|  \gg G$, one component of the doublet is near $\omega_{ml}$ and the other
near $\omega-\omega_{ln}$ . With decreasing $|\Omega|$, the shifted component moves
towards the unshifted one, the latter shifts in the same direction, and the rate of
motion $(M_{1,2})$ is larger for the component that is farthest from $\omega_{ml}$.
The distance between the components is $\sqrt {\Omega^2+4G^2}$. When $\Omega = 0$,
the splitting is symmetrical, and the distance between components is minimal $(2G)$.
Further change of $\Omega$ brings the previously-shifted component closer to
$\omega_{ml}$, and moves the previously-unshifted component at an increased rate.
{\it In addition to the shift of the lines, their Doppler widths also change in
accordance with the memory factors $M_{1,2}$ and the observation direction. The
widths have minimal values $|k_\mu-M_{1,2}k|{\rm\bar v}$} along the direction $\theta
= 0$, {\it and a maximal value $(k_\mu+M_{1,2}k){\rm\bar v}$} in the opposite
direction.

{\it We recall that the use of the obtained results for an analysis of other
two-photon processes implies a reversal of the signs of $\Omega$ and $\Omega_\mu$ in
accordance with whether a particular photon is absorbed or emitted. In the case of
two-quantum luminescence and absorption, the quantities $\pm (\Omega+\Omega_\mu)$ are
involved. Therefore, unlike Raman scattering, the minimum of the Doppler width will
be reached at $\theta=\pi$. The magnitude of the narrowing will be the same as
before.}

\section{DOPPLER BROADENING OF RESONANCE-FLUORESCENCE LINE ON EXCITED LEVELS}

\par
The results of Sec. 2 allow us to explain the correlation and frequency properties of
the radiation also {\it in the case of a transition between the levels that interact
with the strong field $(m \rightarrow n)$}. Within the framework of second-order
perturbation theory, the radiation power $\omega_\mu$ is determined directly by
formula (3.2) in which we put $k = k_\mu$ and $\gamma_l = \gamma_n$. In the angle
interval $\theta \le \gamma_n/k {\rm\bar v}$, the second term will have a dispersion
form with width $2\gamma_n$. Thus, the main conclusions of Sec. 3 concerning the
width anisotropy, the line shift, etc. apply also to resonance fluorescence
\footnote{Resonance fluorescence is usually considered for the case when the lower
level is the ground level $(\gamma_n = 0)$, and transitions from $m$ are allowed only
to $n$. Neither premise is satisfied in our problem. }.

In the case of a strong field $G$, a singularity of the transition is the need for
taking into account the field disturbance of both equations-both the upper and the
lower. As a result, the formula for ${w}_\mu$, in the case of a strong field, is more
complicated than (2.6). For our purposes it suffices, however, to use the general
conclusions of Sec. 2. From formula (2.3) it is easy to conclude that ${w}_\mu$ will
consist of four transitions between two sublevels of the upper state and two
sublevels of the lower state. Each of these transitions contributes its own resonant
term:

{\fs$$
[2\alpha_1^{'}+i(\Omega_\mu-\Omega-{\bf qv})]^{-1}, \quad
[\Gamma+i(\Omega_\mu-{\bf k}_\mu{\bf v}-2\alpha_1^{''})]^{-1}
$$
$$
[2\alpha_2^{'}+i(\Omega_\mu-\Omega-{\bf qv})]^{-1}, \quad
[\Gamma+i(\Omega_\mu-{\bf k}_\mu{\bf v}-2\alpha_2^{''})]^{-1}
\eqno(5.1)
$$}

\noindent In the general case these terms differ in the positions of their maxima (as
functions of $\Omega_\mu$), in their widths, and in the coefficients with which they
enter in ${w}_\mu$. We consider the simplest and most striking case when $G \gg
\Omega, k{\rm\bar v}, \Gamma, \gamma_l$. The amplitudes of all the substates are the
same here, and the fact (5.1) enter in ${w}_\mu$ with equal weights. Further, formula
(2.8) is valid for $\alpha_1$ and $\alpha_2$, and consequently $\langle {w}_\mu
\rangle$ is given by

{\fs$$ \langle { w}_\mu \rangle \propto \langle [\Gamma+i(\Omega_\mu-\Omega-2G-{\bf
qv})]^{-1}+ 2[\Gamma+i(\Omega_\mu-\Omega-{\bf qv})]^{-1}+ $$ $$+
[\Gamma+i(\Omega_\mu-\Omega+2G-{\bf qv})]^{-1} \rangle, \eqno(5.2) $$ $$ {\bf q}={\bf
k}_\mu-{\bf k}, \quad q=2k\sin (\theta/2). $$}

\noindent The position of the maximum of one of the terms coincides with the
frequency $\omega$ of the strong field, and the two others are shifted to the
points $\omega \pm 2G$. The widths of all the components of the triplet have
identical angular characteristics. At large observation angles, the
contours of the lines have a Gaussian form with width $q{\rm\bar v}$.
Inside the cone $\theta<\Gamma/k{\rm\bar v}$, the lines have
a dispersion form with natural width $\Gamma$.

It is of interest to trace the connection between the components of the
triplet (5.2) of the resonant fluorescence with the lines of the stepwise
transition or Rayleigh scattering. If we successively increase the deviation
from resonance, say in the direction of positive $\Omega$, then the component
in (5.2) of frequency $\Omega-2G$ will shift towards $\omega_{mn}$
change into a stepwise-transition line.
The frequency of the central component will increase and
coincide at all time with $\omega$, yielding a Rayleigh-scattering line. Finally,
the third component will move away from $\omega_{mn}$ at a still larger length,
and its amplitude will become of the order of $G^4/\Omega^4$ (i.e.,
it disappears
in the second approximation of perturbation theory). When $\Omega$ changes in
the opposite direction, the unshifted component, as before, remains at
the frequency of the external field, and the roles of the shifted
components are interchanged.
Thus, the changes of the frequency-correlation properties due to an
external field become manifest in the Doppler broadening of the resonance
fluorescence lines to the same degree as in Raman scattering. In addition,
there appears one more line that does not fit in the
classification of second-order perturbation theory.

\vspace*{5mm}
\rule{30mm}{.4pt}

[1] I.M. Fabelinskiy, Molekulyarnoe rasseyanie sveta
(Molecular Scattering of Light), Nauka, 1965 [Consultants Bureau, 1968].

[2] W. Heitler, The Quantum Theory of Radiation, Oxford, 1954.

[3] S.G. Rautian and I.I. Sobel'man, Zh. Eksp. Teor. Fiz. 44, 934 (1963)
[Sov. Phys.-JETP 17, 635 (1963)].

[4] G.E. Notkin, S.G. Rautian, and A.A. Feoktistov, ibid. 52, 1673 (1967) [25, 1112 (1967)].

[5] S.G. Rautian, Trudy FIAN 43, 3 (1968).

[6] S.H. Autler and C.H. Townes, Phys. Rev. 100, 703 (1955).

[7] V.M. Kontorovich and A.M. Prokhorov, Zh. Eksp. Teor. Fiz. 33, 1428 (1957)
[Sov. Phys.-JETP 6, 1100 (1958)].

[8] S.G. Rautian and I.I. Sobel'man, ibid. 41, 456 (1961) [14 328 (1962)].

[9] A.M. Bonch-Bruevich and V.A. Khodovoi, Usp. Fiz. Nauk 93, 71 (1967)
[Sov. Phys.-Uspekhi 10, 637 (1968)].

\vspace*{10mm}
Translated by J. G. Adashko 51

\end{multicols}
\end{document}